\documentclass[sn-mathphys, a4paper]{sn-jnl}

\usepackage[utf8]{inputenc} %
\usepackage[T1]{fontenc}
\usepackage{graphicx}
\usepackage{natbib}
\usepackage{acronym} 

\usepackage{float} 
\usepackage{comment}
\usepackage{amsmath,amssymb} 
\usepackage{xcolor}

\usepackage{hyperref}
\usepackage[noabbrev]{cleveref} 
\usepackage{comment}
\usepackage{caption}
\usepackage{subcaption}
\usepackage{xcolor}
\def\ie{{\em i.e. }} 
 
\acrodef{3d}[3D]{3-dimensional} 
\acrodef{3c-3d}[3C-3D]{3-component--3-dimensional}
\acrodef{piv}[PIV]{particle image velocimetry}
\acrodef{dns}[DNS]{direct numerical simulation}
\acrodefplural{dns}[DNSs]{direct numerical simulations}
\acrodef{les}[LES]{large eddy simulation}
\acrodef{cfd}[CFD]{computational fluid dynamics}
\acrodef{tsf}[TSF]{turbulent shear flow}
\acrodefplural{tsf}[TSFs]{turbulent shear Flows}
\acrodef{tbl}[TBL]{turbulent boundary layer}
\acrodefplural{tbl}[TBLs]{turbulent boundary layers}
\acrodef{ai}[AI]{artificial intelligence}
\acrodef{ml}[ML]{machine learning}
\acrodef{sciml}[SciML]{scientific machine learning}
\acrodef{rans}[RANS]{Reynolds-averaged Navier--Stokes}
\acrodef{les}[LES]{large eddy simulation}
\acrodefplural{les}[LESs]{large eddy simulations}
\acrodef{dof}[DoF]{degrees of freedom}
\acrodefplural{DoF}[DoFs]{degrees of freedom}
\acrodef{rom}[ROM]{reduced-order model}
\acrodefplural{rom}[ROMs]{reduced-order models}
\acrodef{da}[DA]{data assimilation}
\acrodef{vda}[VDA]{variational data assimilation}
\acrodef{vgt}[VGT]{velocity gradient tensor}
\acrodef{tke}[TKE]{turbulent kinetic energy}
\acrodef{pod}[POD]{proper orthogonal decomposition}
\acrodef{ltrac}[LTRAC]{Laboratory for Turbulence Research in Aerospace and Combustion}
\acrodef{hpc}[HPC]{High Performance Computing}
\acrodef{ode}[ODE]{ordinary differential equation}
\acrodefplural{ode}[ODEs]{ordinary differential equations}
\acrodef{pde}[PDE]{partial differential equation}
\acrodefplural{pde}[PDEs]{partial differential equations}
\acrodef{zpg}[ZPG]{zero-pressure-gradient}
\acrodef{cfd}[CFD]{computational fluid dynamics}
\acrodef{camf}[CAMF]{Coupled Atmosphere Wildland Fire Environment}
\acrodef{ttbl}[TTBL]{turbulent thermal boundary layer}
\acrodef{tbl}[TBL]{turbulent boundary layer}
\acrodef{shpc}[S-HPC]{scientific high-performance computing}
\acrodef{rs-rt-3d-bftv}[RS-3D-BFTV]{{\bf remote-sensing 3D bushfire temperature and velocimetry}}
\acrodef{rtpibp}[RT-BFP]{{\bf real-time bushfire predictor} }
\acrodef{nn}[NN]{neural network}
\acrodef{dnn}[DNN]{deep neural network}
\acrodefplural{dnn}[DNNs]{deep neural networks}
\acrodef{pinn}[PINN]{physics-informed neural network}
\acrodefplural{pinn}[PINNs]{physics-informed neural networks}
\acrodef{pidl}[PiDL]{physics-informed deep learning}
\acrodef{cnn}[CNN]{convolutional neural network}
\acrodefplural{cnn}[CNNs]{convolutional neural networks}
\acrodef{rnn}[RNN]{recurrent neural network}
\acrodefplural{rcnn}[RNNs]{recurrent neural networks}
\acrodef{gan}[GAN]{generative adversarial network}
\acrodefplural{gan}[GANs]{generative adversarial networks}
\acrodef{ae}[AE]{autoencode}
\acrodef{vae}[VAE]{variational autoencoder}
\acrodefplural{vae}[VAEs]{variational autoencoders}
\acrodef{ae}[AE]{autoencoder}
\acrodefplural{ae}[AEs]{autoencoders}
\acrodef{dl}[DL]{deep learning}
\acrodef{ml}[ML]{machine learning}
\acrodef{uq}[UQ]{uncertainty quantification}
\acrodef{dp}[DP]{Discovery Project}
\acrodef{uav}[UAV]{unmanned aerial vehicle}
\acrodefplural{uav}[UAVs]{unmanned aerial vehicles}
\acrodef{rf}[RF]{Research Fellow}
\acrodefplural{rf}[RFs]{Research Fellows}
\acrodef{hdr}[HDR]{Higher Degree Researcher}
\acrodef{fem}[FEM]{Finite Element Method}
\acrodefplural{hdr}[HDRs]{Higher Degree Researchers}

\begin{document}

\title{Direct numerical simulation of a thermal turbulent boundary layer: an analogy to simulate bushfires and a testbed for artificial intelligence remote sensing of bushfire propagation}

\author*{Julio Soria}\email{julio.soria@monash.edu}
\author{Shahram Karami}
\author{Callum Atkinson}
\author{Minghang Li}

\affil{\orgdiv{Laboratory for Turbulence Research in Aerospace \& Combustion (LTRAC), Department of Mechanical and Aerospace Engineering}, \orgname{Monash University}, \orgaddress{\street{(Clayton Campus)}, \city{Melbourne}, \postcode{3800}, \state{Victoria}, \country{Australia}}}

\abstract{
Direct numerical simulation of a \ac{ttbl} can perform the role of an analogy to simulate bushfires that can serve as a testbed for \ac{ai} enhanced remote sensing of bushfire propagation. By solving the Navier-Stokes equations for a turbulent flow, DNS predicts the flow field and allows for a detailed study of the interactions between the turbulent flow and thermal plumes. In addition to potentially providing insights into the complex bushfire behaviour, \ac{dns} can generate synthetic remote sensing data to train \ac{ai} algorithms such as \acp{cnn} and \acp{rcnn}, which can process large amounts of remotely sensed data associated with bushfire. Using the results of \ac{dns} as training data can improve the accuracy of \ac{ai} remote sensing in predicting fire front propagation of bushfires. \ac{dns} can also test the accuracy of the \ac{ai} remote sensing algorithms by generating synthetic remote sensing data that allows their performance assessment and uncertainty quantification in predicting the evolution of a bushfire. The combination of \ac{dns} and \ac{ai} can improve our understanding of bushfire dynamics, develop more accurate prediction models, and aid in bushfire management and mitigation.
}

\maketitle

\section{Introduction}\label{sec:Intro}

\begin{figure}[htbp]
	\centering
	\includegraphics[width=\textwidth]{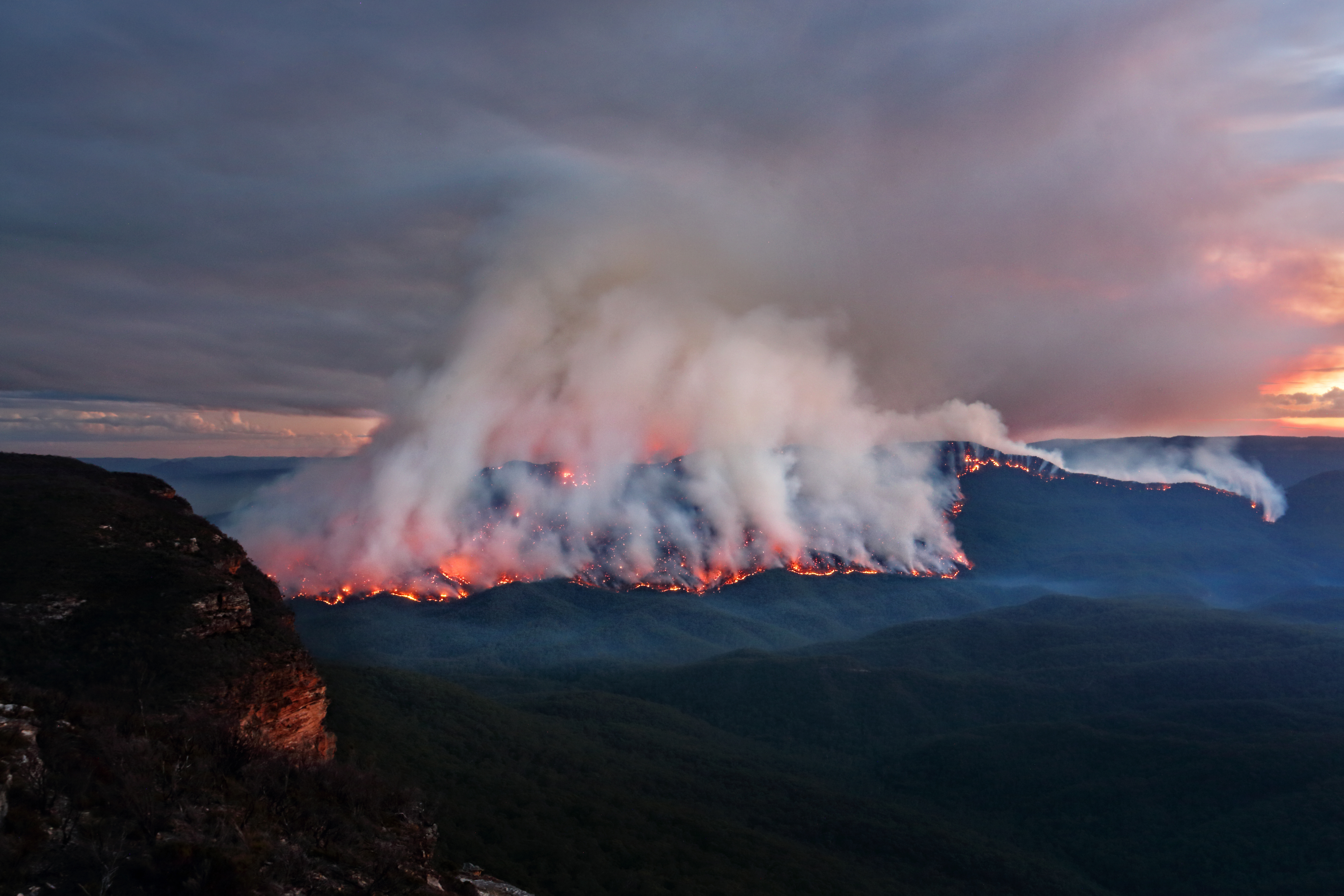}
	\caption{Blue Mountains bushfire (Gospers Mountain, NSW Australia, December 2019) burned more than 510,000 ha (1.26 million acres) making it the largest forest fire ever recorded in Australia \cite{BlueMountaiFire2019}}
    \label{fig:Fig-1}
\end{figure}

Australia, like many other places in the world with a similar climate, experiences frequent bushfires or wildfires, which as was recently experienced, pose a serious threat to the local population, unique Australian wildlife, and natural resources, as illustrated in Fig. \ref{fig:Fig-1} (\cite{BlueMountaiFire2019, Damany-Pearce2022}). These bushfires also contribute to an increase in global CO\textsubscript{2} emissions.

Despite decades of research on bushfires (\cite{nla.cat-vn8555232, CSIRO2020}), there are no verified theories on how they spread that can serve as a basis for accurate prediction of bushfire dynamics. However, such understanding is necessary to develop optimal fire prevention, mitigation, and control strategies to minimise the negative impact on the population and the environment. \ac{cfd} or lower resolution numerical weather prediction models of the atmosphere coupled with empirical (or quasi-empirical) fire spread models, as used in front-tracking approaches, are unsuitable for operational fire spread prediction \cite{CSIRO2020}.

Presently, one approach for remote monitoring of bushfires involves using infrared satellite sensing. However, while this approach provides temperature data, it does not provide information on how and at what rate the fire is moving and behaving within the complex multi-scale, highly dynamic, \ac{tbl} of the atmosphere. As a result, this approach cannot accurately predict how the fire front will spread.  To tackle this problem, there is a requirement to supplement the temperature data obtained from satellites or other aerial platforms like \acp{uav} fitted with infrared cameras with the ability to predict the behaviour and movement of bushfires, including the spreading rate of the fire front. Such a predictive ability would represent a significant improvement in all aspects of the disaster management cycle of bushfires. Therefore, this paper proposes a domain-specific \ac{sciml} methodology that utilises physics-informed \ac{ml} \cite{Baker:2019up, WilcoxICIAM2019, Fukami:2021ka}  based on \ac{dl} to assimilate remotely sensed temperature data and yield the augmented information on bushfire heat release rates and transport properties, such as wind velocity, convective energy transport, and their spatio-temporal evolution. The real-time generation and availability of such augmented bushfire dynamics information has the potential to be a vital new tool in all aspects of the disaster management cycle of bushfires, including prevention, planning, response, and recovery, by providing high-fidelity, low-uncertainty predictions of bushfire propagation that is currently unavailable.
In order to develop, test and undertake \ac{uq} of the proposed \ac{sciml} methodology that will provide the predictive capability of bushfire dynamics and fire front propagation by enhancing remotely sensed data, it is imperative to have high-quality fully resolved and completely charatcerised \ac{ttbl} data to serve as the {\em “ground truth”}. This fully resolved and quantified \ac{ttbl} data of a bushfire is only available via \ac{dns}. However, a realistic bushfire \ac{dns} with real world topographies of which there are a large multitude, with all its multi-physics including combustion is computationally infeasible at the moment, even with the most powerful Exascale supercomputers. Furthermore, the actual parameter space is incredibly large, making \acp{dns} of all possible cases to establish the necessary statistical data for all possible scenarios unfeasible. Therefore, in this paper an analogy to a bushfire is proposed to serve as the {\em “ground truth”} and to be the testbed for \ac{ai} enhanced remote sensing of bushfire propagation. 
 This paper briefly describes the proposed \ac{sciml} methodology that utilises physics-informed \ac{ml} based on \ac{dl} to assimilate remotely sensed temperature data and the approach to its \ac{uq}. This is followed with a description of the \ac{ttbl} \ac{dns} methodology, which permits distributed and quite arbitrary energy sources to be defined and which includes a temperature-dependent heat-release model as a source term in the energy equation \cite{Li2022}. The proposed implementation of the energy source distribution allows biomass fuels, i.e. grass, scrubs, trees, etc., to be modelled as localised energy source with temperature dependent energy release rate depreciating at the same rate as the source term in the energy equation. Finally the results of one \ac{ttbl} \ac{dns}, which serves as the {\em “ground truth”} to develop, test and \ac{uq} any \ac{ai} enhanced remotely sensed data of a bushfire are illustrated.

\section{SciML Based on Physics-informed Deep Learning for Bushfire Predictions}\label{sec:SciML}
Currently, there is significant effort into remote sensing for detection and prediction of bushfires/wildfires, such as the Fire Urgency Estimator in Geosynchronous Orbit (FUEGO) \cite{Pennypacker2013}. However, there is currently no operational nor proposed system which employs \ac{ai} in the form of discipline specific SciML to enhance this type of data and provide bushfire dynamics and reliable fire front propagation rate predictions. The proposed discipline specific \ac{sciml} presented in this paper is planned to employ remote sensing by Satellites, and aerial platforms like UAV, etc. fitted with infrared sensors to acquire 3D temperature field information of bushfires as outlined \cite{Pennypacker2013}. Furthermore the idea here is to couple this 3D temperature field information with \ac{da}/\ac{sciml} to provide information on the heat release rate distribution,  turbulence, transport and fire front spreading rate. This will enable the development of an improved \ac{ai} enhanced predictive model to forecast bushfire spreading rates. 

\begin{figure}[htbp]
	\centering
	\includegraphics[width=\textwidth]{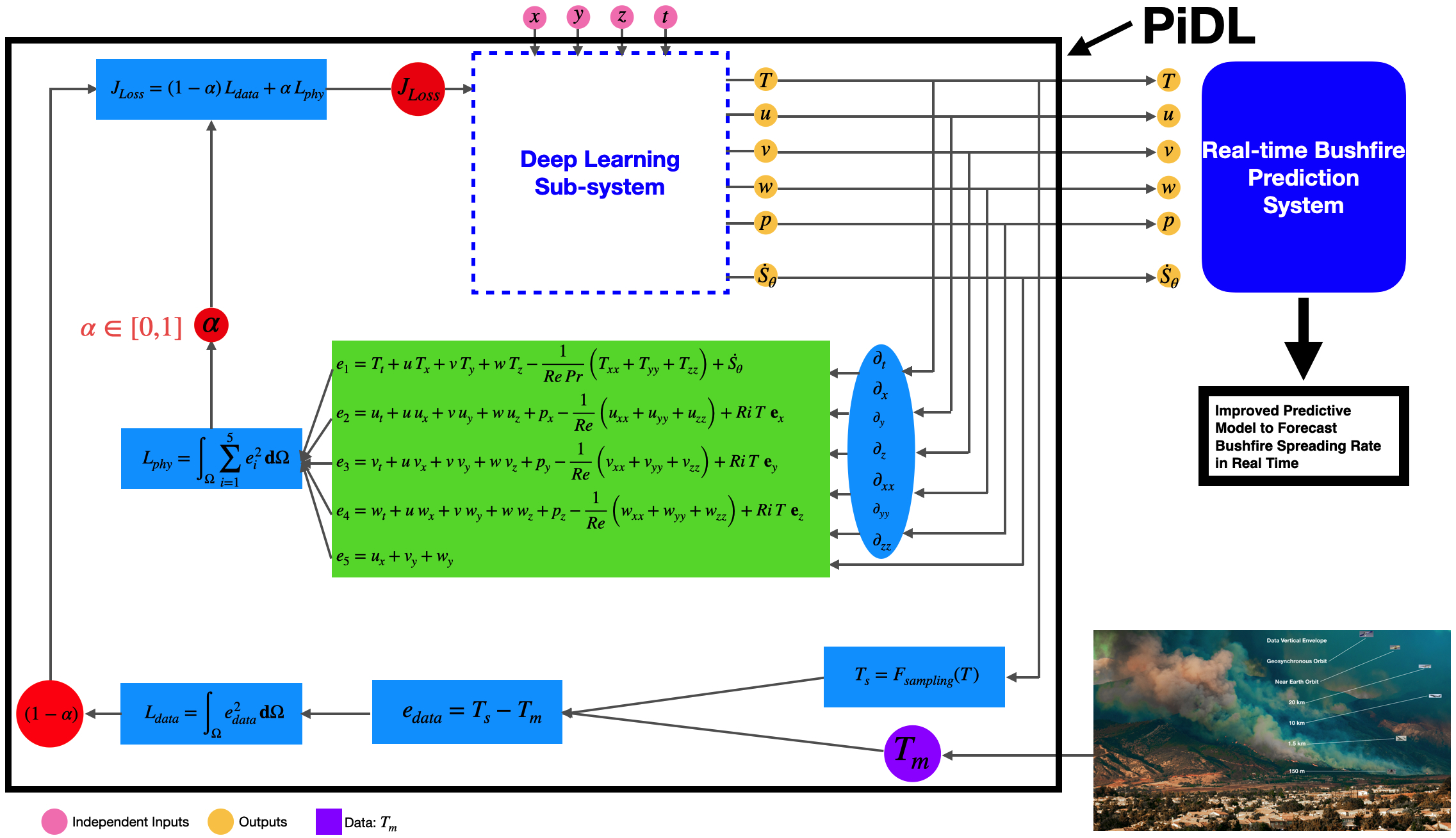}
	\caption{Proposed \ac{sciml} for bushfire predictions based on \ac{pidl}. The components of the system within the black rectangular box represent the \ac{pidl}. The inputs to the \ac{pidl} is the independent 4-dimensional vector space of space and time $(x, y, z, t)$ and the sampled 3D temperature data $T_m$, which is here pictorially represented from data that is obtainable from the FUEGO vertical envelope \cite{Pennypacker2013}. The \ac{ai} learned output of the \ac{pidl} is the \ac{ttbl} turbulent velocity vector field, the 3D turbulent temperature field and the source term which represents the heat release due to burning of the biomass during the bushfire. All outputs are provided by the PiDL as functions of 4-dimensional vector space of space and time $(x, y, z, t)$.}
    \label{fig:Fig-2}
\end{figure}

Figure \ref{fig:Fig-2} shows the proposed \ac{sciml} based on \ac{pidl} for bushfire predictions. The inputs to the \ac{pidl} is the 4-dimensional vector space of three-dimensional (3D) space and time, $(x, y, z, t)$, which can be defined on an arbitrary three-dimensional unstructured grid, and the sampled 3D temperature data, $T_m(x_m, y_m, z_m,t_m)$, which can come from a range of sources fitted with infrared sensors ranging from low orbit satellites to fixed wing aircraft to UAVs or drones and serves as the training data set for the \ac{pidl}. The \ac{pidl} uses this data to both post-dict (via training) and predict the \ac{ttbl}  turbulent velocity vector field, the turbulent pressure and temperature fields, as well as the source term, which represents the heat release rate due to burning of the biomass during the bushfire, \ie $\left(u(x,y,z,t), v(x,y,z,t), w(x,y,z,t)\right), p(x,y,z,t), T(x,y,z,t) \mbox{ and } \dot{S}_\theta(x,y,z,t)$.

The \ac{pidl} is contained entirely within the black rectangular box in Fig. \ref{fig:Fig-2} with only the independent 4-dimensional vector space of (3D) space and time, the sampled 3D temperature data and the turbulent velocity vector, pressure and temperature fields as well as source term field crossing its boundaries.
Within the \ac{pidl} is the \ac{dl} sub-system, which can be a fully connected \ac{nn}, \ac{cnn} or a \ac{rnn}, etc., the disciple specific physics-information in the form of the incompressible Navier-Stokes equations, the temperature energy equation, which are both liked via a Boussinesq approximation and written in terms of residual $e_i, i \in [1,2,3,4,5]$, for the 5 governing equations. The temperature output from the \ac{dl} sub-system, $T(x,y,z,t)$, is spatially and temporally sampled to be spatially and temporally coincident with the sampled 3D temperature data $T_m(x_m, y_m, z_m, t_m)$ resulting from the application of the sampling operator $F_{sampling}(T)$ resulting in $T_s(x_m, y_m, z_m, t_m)$, which is compared to $T_m(x_m, y_m, z_m, t_m)$ and yields a residual $e_{data}$.

As shown in Fig. \ref{fig:Fig-2} the $e_i’s$ are combined and integrated over space and time to yield an error norm pertaining to the physics information $L_{phy}$, while the $e_{data}$ is also integrated over its space and time to yield an error norm $L_{data}$. These two error norms are combined into a cost function $J_{Loss}$ using complementary weights given by the \ac{pidl} parameter $\alpha \in [0,1]$ , which is minimised to train the \ac{dl} sub-system. 

Note that $\alpha = 1$ corresponds to a \ac{pidl} that is purely a \ac{pde} \ac{dl} solver. Although highly inefficient compared to traditional numerical methods and it would have to be augmented with additional physics information in the form of boundary and initial conditions, otherwise it would produce nonsensical results.

An $\alpha = 0$ corresponds to purely machine learning using the \ac{dl} system, but no physics information. This could yield temperature interpolation, but no sensible turbulent fluid velocity and pressure field and heat source information and since there is no physical constraints on this version of the \ac{pidl} it could also produce nonsensical temperature interpolation results. Hence, away from these two limits of $\alpha$, both physics information and measured temperature data contribute to the training and predictions of the \ac{pidl} with its performance therefore dependant on $\alpha$.

\subsection{“Ground Truth” Data and Uncertainty Quantification of \ac{pidl}}

\begin{figure}[htbp]
	\centering
	\includegraphics[width=\textwidth]{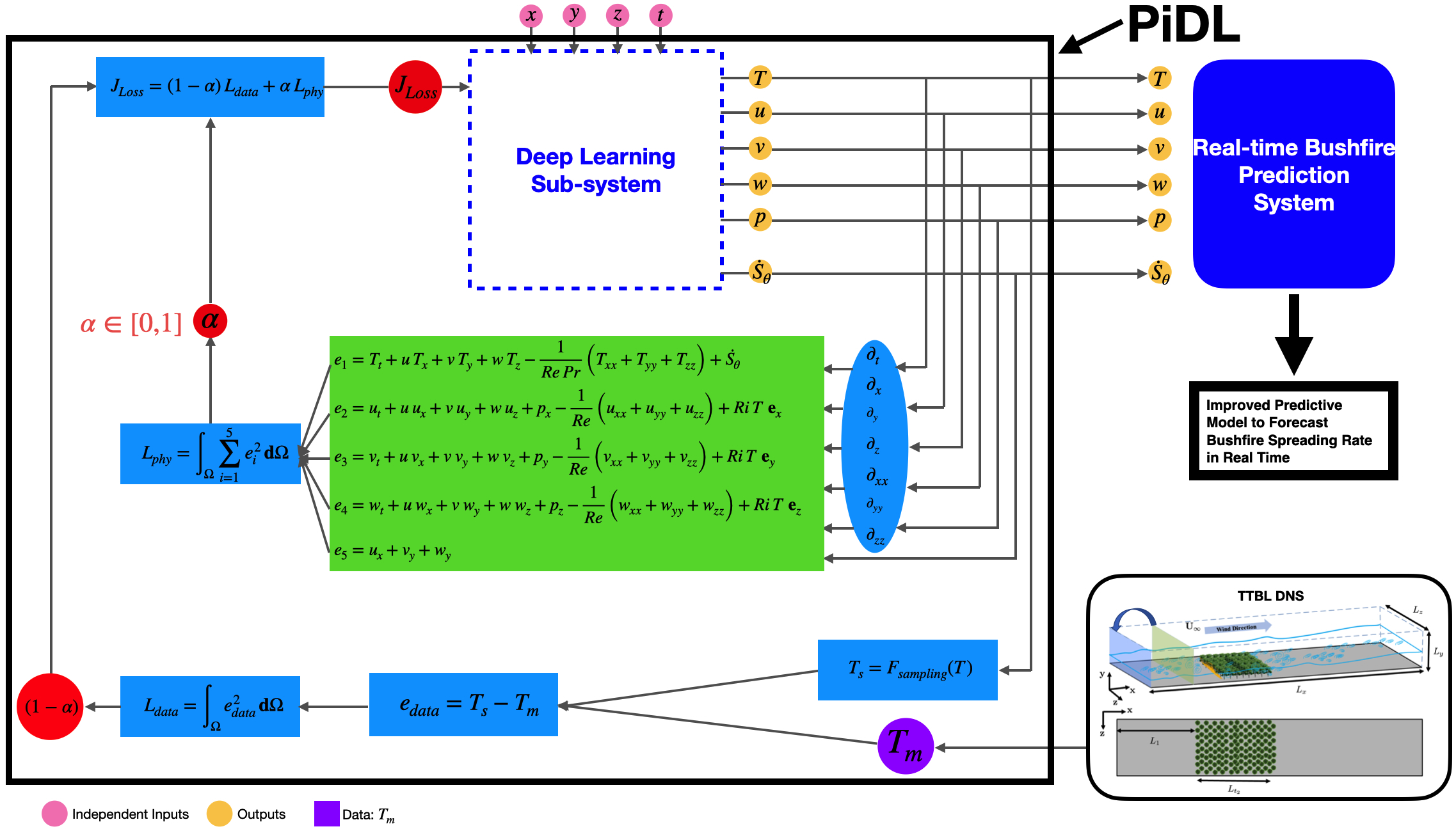}
	\caption{Development, testing and \ac{uq} of proposed \ac{sciml} based on \ac{pidl} for bushfire predictions using {\em “ground truth”} \ac{ttbl} \ac{dns} temperature data. The approach to \ac{uq} of the \ac{pidl} is illustrated in Fig. \ref{fig:Fig-4}. The \ac{ttbl} \ac{dns} fully resolved temperature data $T_e(\mathbf{x}, t)$ is downsampled at random locations and a given sparsity with measurement noise introduced to simulate measurement uncertainty, which yield the sampled 3D temperature data $T_m(\mathbf{x}_m, t_m)$.}
    \label{fig:Fig-3}
\end{figure}

In order to develop the details of the \ac{pidl}, \ie the specific \ac{dl} sub-system, and test it, \ie determine the sensitivity of the \ac{pidl} output with respect to $\alpha$ and/or find its optimal value and \ac{uq} of the \ac{pidl}, {\em “ground truth”} data of every aspect is invaluable. However, this type of data is impossible to acquire from real bushfires and even limited experimental laboratory “bush” fires due to the current complete absence of full field 3D simultaneous velocity vector and temperature field measurement techniques to provide any reliable measurements in the highly challenging environment of a bushfire. Nevertheless, \ac{dns} of a \ac{ttbl} with distributed and quite arbitrary energy sources with a well-defined temperature-dependent heat-release rate model as a source term in the energy equation \cite{Li2022}, serving as an analog to a bushfire provides the full highly resolved turbulent velocity vector field, the pressure and temperature field and the heat release source term as the necessary {\em “ground truth”} data as shown in Fig. \ref{fig:Fig-3}.

The output from the \ac{pidl} at the same spatial resolution as the \ac{ttbl} \ac{dns}: $\mathbf{u}(\mathbf{x}, t)$, $p(\mathbf{x},t)$, $T(\mathbf{x}, t)$ and $\dot{S}_\theta(\mathbf{x}, t)$ is then used with the {\em “ground truth”} \ac{ttbl} \ac{dns} data:  $\mathbf{u}_e(\mathbf{x}, t)$, $p_e(\mathbf{x}, t)$, $T_e(\mathbf{x}, t)$ and $\dot{S}_{\theta_e}(\mathbf{x}, t)$ to \ac{uq} the \ac{pidl} with respect to bias error, uncertainty, time horizon predictability, etc. as a function of sampling spacial resolution, sparsity and measurement uncertainty introduced via noise in $T_m(\mathbf{x}_m, t_m)$.

\begin{figure}[htbp]
	\centering
	\includegraphics[width=\textwidth]{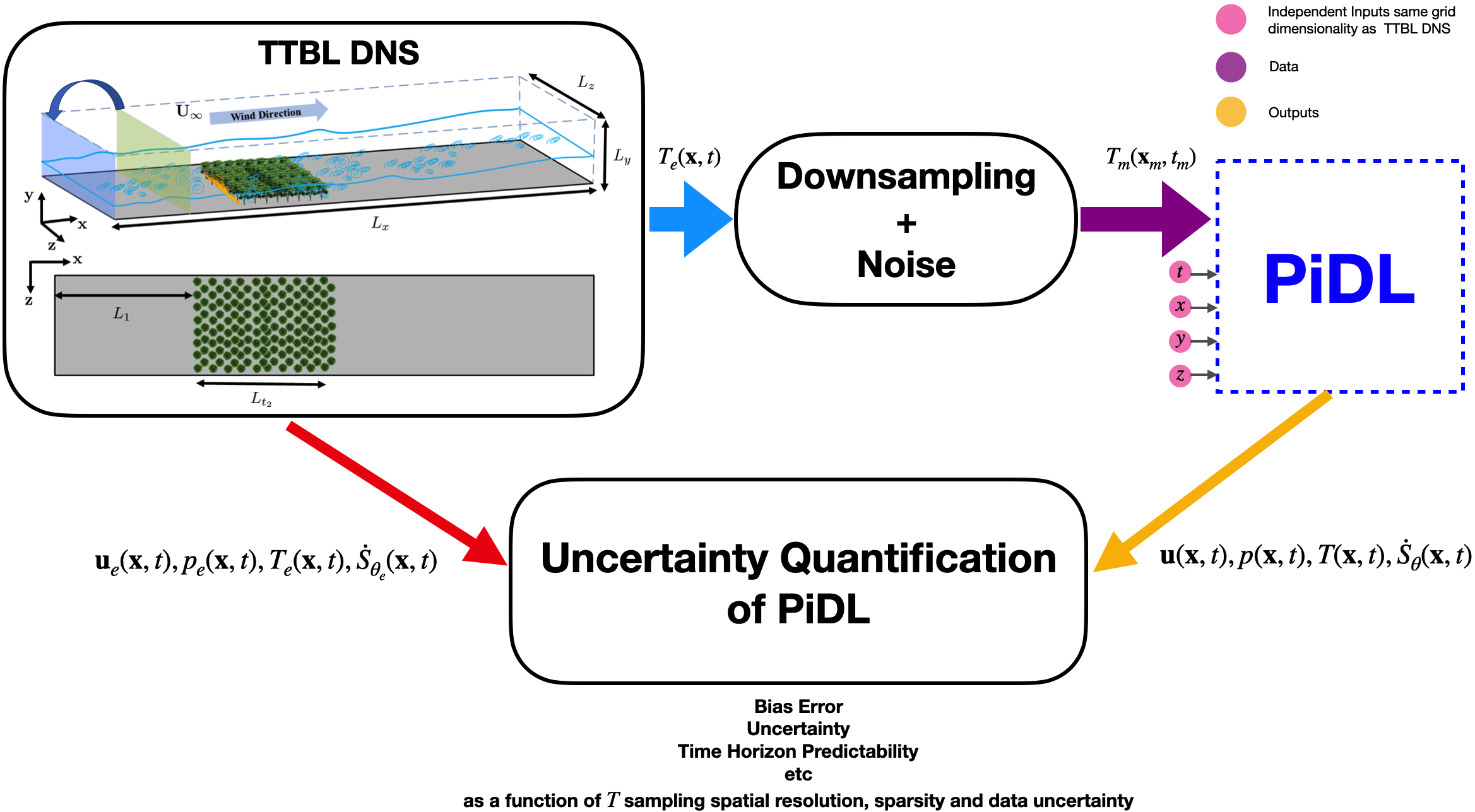}
	\caption{System diagram for \ac{uq} of the proposed \ac{sciml} based on \ac{pidl} for bushfire predictions using {\em “ground truth”} \ac{ttbl} \ac{dns} data. Variables with the subscript "$e$" indicates \ac{ttbl} \ac{dns} data, variables with the subscript "$m$" indicates simulated measured temperature data, while variables without any subscript indicate output variables from the \ac{pidl}.}
    \label{fig:Fig-4}
\end{figure}

\section{Direct Numerical Simulation of a Thermal Turbulent Boundary Layer - An Analogy to Simulate Bushfires}\label{sec:dns}

Given the enormous parameter space necessary to represent all bushfire scenarios and fuel load characteristics, it is clearly not possible to develop a universal mathematical model of bushfire and fire propagation that provides pertinent information in real-time and with high confidence, which is essential for bushfire management. This will in fact be the work of the fully implemented, tested and \ac{uq} \ac{pidl}, which it is hypothesised, will learn from the field data all these necessary details. 

Instead, \ac{dns} is used to accurately simulate a canonical \ac{ttbl} case, which is an analogy to a bushfire as a first step by including a fully resolved \ac{zpg}-\ac{tbl} with the bushfire modelled as a distribution of localised heat sources with a temperature dependent energy release rate  \cite{Li2022}.

\subsection{Theoretical Framework of the TTBL DNS}

The governing equations for the \ac{ttbl} are the incompressible Navier-Stokes equations with the temperature-energy equation, which includes an energy source term to model the heat release rate due to the burning of the biomass representing the bush, etc. The Navier-Stokes equations are coupled to the temperature-energy equation via a Boussinesq approximation \cite{Tritton1977} and are given by:

\begin{eqnarray}
\frac{\partial u_i}{\partial x_i} &=& 0  \label{eq:cont} \\
\frac{\partial u_{i} }{\partial t } +\frac{\partial u_{i} u_{j} }{\partial x_{j}} &=& -\frac{\partial p }{\partial x_{i} } +\frac{1}{\mathrm{Re}}\frac{\partial^2 u_{k} }{\partial x^2_{k}} + \frac{\kappa_i}{\mathrm{Fr_o}^2}\theta  \label{eq:mom} \\
\frac{\partial \theta }{\partial t } + \frac{\partial u_{j} \theta}{\partial x_{j} } &=& \frac{1}{Re\, Pr} \frac{\partial^2 \theta }{\partial x^2_{i}}+ \dot{S}_{\theta}, \label{eq:temp}
\end{eqnarray}

where Eq. \ref{eq:cont} represents the continuity equation, a statement of conservation of mass, Eq.  \ref{eq:mom} represents the 3 momentum equations, which are a statement of conservation of linear momentum and Eq. \ref{eq:temp}  is the temperature-energy equation, which is a statement of conservation of energy. The last term on the right hand side of Eq.  \ref{eq:mom} is the Boussinesq approximation with $\kappa_i$ a vector indicating the direction of the gravitational field which in this case is $[0,-1,0]$, while the last term on the right-hand side of Eq. \ref{eq:temp}, $\dot{S}_{\theta}$, represents the source term, which provides the heat release rate from the burning of biomass sources. This term needs to be modelled.

Equations \ref{eq:cont} - \ref{eq:temp} are written in non-dimensional form with respect to the non-dimensional spatial vector $x_j$, velocity vector $u_i$, pressure $p$ and temperature $\theta$, defined with respect to its dimensional starred independent and dependent variables, fluid  properties and parameters as:

\begin{equation}
\begin{array}{l}
x_j = \frac{x_j^*}{L}; \;\; t = \frac{t^*\, U_\infty^*}{L}; \;\; \\[2mm]
u_j = \frac{u_j^*}{U_\infty^*};  \;\; p = \frac{p^*}{\rho_\infty^*\, U_\infty^{*2}}; \;\; \theta = \frac{T^* - T_\infty^*}{T_{ad}^* - T_\infty^*}; \;\; \zeta_0 = \frac{T^*}{T_{ad}^* - T_\infty^*};\\[2mm]
\rho = \frac{\rho^*}{\rho_\infty^*}; \;\; \mu = \frac{\mu^*}{\mu_\infty^*};\;\;  Re = \frac{\rho_\infty^*\, U_\infty^*\, L}{\mu_\infty^*}; \;\; Pr = \frac{\nu_\infty^*}{\alpha_\infty^*}; \;\; Fr = \frac{U_\infty^*}{\sqrt{g\, L}}; \;\;\; Fr_0 = \zeta_0^{1/2}\, Fr
\end{array}
\label{eq:non-dim}
\end{equation}

where the characteristic quantities:  $L$ = characteristic length, $U_\infty^*$ = characteristic free-stream velocity, $\rho_\infty^*$ = characteristic density, $T_\infty^*$  = characteristic free-stream temperature, $\mu_\infty^*$ = characteristic dynamic viscosity, $\nu_\infty^* = \frac{\mu_\infty^*}{\rho_\infty^*}$ = characteristic kinematic viscosity, $\alpha_\infty^*$ = characteristic thermal diffusivity,  $g$ = acceleration due to gravity and  $T_{ad}^*$ = characteristic adiabatic flame temperature. The non-dimensional parameters $Re$ = Reynolds number, $Pr$ = Prandtl number and $Fr$ = Froude number.

\subsection{Model for the Heat Release Rate of the Burning of Biomass}

The source term in Eq. \ref{eq:temp}  which accounts for the heat release rate of the burning of biomass in bushfires needs to be modelled. For simplicity, an exothermic reaction of the type $F \longrightarrow P$, where $F$ is the unburned biomass and $P$ is the product, is assumed. Furthermore, it is assumed that the biomass and product have the same constant heat capacity, molecular weight and molecular diffusion coefficient,. The fuel mass fraction and temperature transport equation can then be written as:

\begin{equation}
\frac{\partial Y^*_{F} }{\partial t^* } +\frac{\partial u^*_{j} Y^*_{F} }{\partial x^*_{j} } =\frac{\partial }{\partial x^*_{j} } \left(D^*\frac{\partial Y^*_{F} }{\partial x^*_{j} }  \right) - \dot{\omega }^*_{F}
\label{eq:FuelMassFraction}
\end{equation}

and

\begin{equation}
\frac{\partial T^*}{\partial t^* } +\frac{\partial u^*_{j}  T^* }{\partial x^*_{j} } = \frac{\partial }{\partial x^*_{j} } \left(\frac{\lambda^*}{c^*_p}\frac{\partial  T^* }{\partial x^*_{j} }  \right) + \frac{Q^*}{c^*_p} \dot{\omega }^*_{F}
\label{eq:T-Transport}
\end{equation}

respectively, where $D^*$ = molecular diffusion coefficient, $\lambda^*$ = heat diffusion coefficient, $Q^*$ = heat released per unit mass of fuel and $\dot{\omega}_F$ = fuel reaction rate. The fuel reaction rate depends on temperature with the relationship

\begin{equation}
\dot{\omega}^*_{F} = B Y^*_{F} T^{*^{\beta_1}} \exp [\frac{-T_{a}}{T^*} ]
\label{eq:ReactionRate}
\end{equation}

where $B$ is a constant, $\beta_1$  is temperature exponent and $T_a$ denotes the constant activation temperature for the Arrhenius reaction, \ie the ratio of the activation energy to the universal gas constant. It should be noted that the density of the gas mixture has been assumed to be constant.

The non-dimensional form of these equations using $\theta$ as the non-dimensionalised temperature and $Y = Y^*_{F}/Y_{F,0}$ as the non-dimensionalised mass fraction, where $Y_{F,0}$ = mass fraction of the fuel in the mixture and using $c^*_pT_{\infty}^{*}+Q^*Y_{F,0} = c^*_p T^{*}_{ad}$ , leads to the non-dimensional form of Eqs. \ref{eq:FuelMassFraction} and \ref{eq:T-Transport}:

\begin{equation}
\frac{\partial Y }{\partial t } +\frac{\partial u_{j} Y }{\partial x_{j} } = \frac{1}{Re\, Pr} \frac{\partial^2 Y }{\partial x^2_{i}}- \frac{\dot{\omega }_{F}}{Y_{F,0}}
\label{eq:NonD-FuelMassFraction}
\end{equation}

and

\begin{equation}
\frac{\partial \theta}{\partial t} +\frac{\partial u_{j}  \theta }{\partial x_{j} } = \frac{1}{Re\, Sc} \frac{\partial^2 \theta }{\partial x^2_{i}} + \frac{\dot{\omega }_{F}}{Y_{F,0}},
\label{eq:NonD-T-Transport}
\end{equation}

respectively. Here $Sc = \frac{\nu^*}{D^*}$ = Schmidt number. If the Lewis number, which is defined as $Le = \frac{Sc}{Pr} =1 $, then adding these two equations leads to a transport equation for $\theta + Y$ with no source term. Considering $Y$ and $\theta$ has a value of 1 and 0, respectively in the unburned mixture and 0 and 1, respectively in the burned mixture, the solution this resulting equation is $\theta + Y = 1$. Hence, the temperature-energy transport equation, Eq. \ref{eq:temp} is the only equation that needs to be solved in addition to the mass and momentum equations Eq. \ref{eq:cont} - \ref{eq:mom}. If the non-dimensionalised temperature defined in Eq. \ref{eq:non-dim} is used for non-dimensionalisation, then the non-dimensionalised temperature-energy equation is given by Eq. \ref{eq:temp} with the source term, $\dot{S}_{\theta} $ appearing in this equation given by

\begin{equation}
\dot{S}_{\theta} = C  (\zeta_o + \theta)^{\beta_1} (1- \theta)  \exp [\frac{-\beta(1-\theta)}{1-\alpha(1-\theta)} ],
\label{eq:source}
\end{equation}

where $\alpha=(T^*_{ad} - T^*_{\infty})/T^*_{ad}$,  $\beta= \alpha T_a/T^*_{ad}$ and $C$ depends on $c^*_p$ = heat capacity at constant pressure, $B$, $Q$  and $Y_{F,0}$. It is not an unreasonable assumption to consider that $Y_{F,0}$ is a constant where any $c^*_p$ variation with temperature can be modelled by adjusting ${\beta_1}$. Hence, the source term can be simplified to read

\begin{equation}
\dot{S}_{\theta} = A\, (1- \theta)\,  \exp \left[\frac{-\beta\,(1-\theta)}{1-\alpha\,(1-\theta)} \right],
\label{eq:SimplifiedSource}
\end{equation}

where $A$ is a constant.

This approach is the centre of many early development in theoretical combustion (\cite{Clavin:1979,Bychkov:2000}). Despite its restrictive assumptions, especially at the chemistry level, this source term model preserves many features, including non-linear heat release and variable temperature.

The heat release parameter $\alpha$ in Eq. \ref{eq:SimplifiedSource} depends on the maximum flame temperature. A recent systematic experimental measurement of the flame temperature of fires in dry eucalyptus forest by \cite{Wotton:2012} reported the maximum temperature from approximately 700$^{\circ}C$ to 1200$^{\circ}C$. This maximum flame temperature leads to $\alpha$ values between 0.69 and 0.8, which can be considered a hyper-parameter.

\begin{figure}[htbp]
	\centering
	\includegraphics[width=\textwidth]{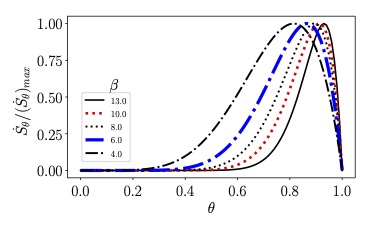}
	\caption{Variations of the normalised heat release rate source term $\dot{S}_{\theta}/(\dot{S}_{\theta})_{max}$ for $\alpha = 0.75$ and various values of $\beta$ as a function of the non-dimensionalised temperature $\theta$.}
    \label{fig:Fig-5}
\end{figure}

Another parameter is the Zeldovich number $\beta$ in Eq. \ref{eq:SimplifiedSource}, which is difficult to obtain for the proposed simple heat release model. However, it is typically below 10 and no more than about 15 in hydrocarbon flames at atmospheric pressure (\cite{Peters:1997, Dold:2002}). Considering a heterogeneous reaction that occurs, for instance, at the gas-to-solid interface with the main source of fuel being solid carbon, the Zeldovich number can be considered as approximately 8.0. Variations of the normalised heat source term $\dot{S}_{\theta}/(\dot{S}_{\theta})_{max}$ for $\alpha = 0.75$ and various values of $\beta$ are presented in Fig. \ref{fig:Fig-5}.

\subsection{Numerical Method}

The high-fidelity simulation of \ac{ttbl} is conducted using a modified version of a hybrid parallel MPI/OpenMP \ac{tbl} \ac{dns} code \cite{Simens:2009mz,Borrell:2012ks,Kitsios-etal-IJHFF:2016,Kitsios:2017eu}. The code solves the three-dimensional incompressible Navier-Stokes equations in a three-dimensional rectangular volume. The three flow directions are streamwise,  (flow direction), wall-normal,  and spanwise  (cross-flow). The numerical simulation code uses the fractional-step method of \cite{Harlow:1965} to solve the governing equations for the velocity, temperature and pressure fields. Fourier decomposition is used in the periodic spanwise direction with 2/3-dealising, with compact finite difference \cite{Lele:1992} used in the streamwise and wall-normal directions. The equations are stepped forward in time using a modified three sub-step Runge-Kutta scheme \cite{Simens:2009mz}. The bottom surface is a flat plate with a no-slip boundary condition. The thermal boundary condition can be specified as a constant temperature or a constant heat flux. The former is used in the \ac{ttbl} \ac{dns} results presented here.

\subsection{Results}

\begin{figure}[htbp]
	\centering
	\includegraphics[width=\textwidth]{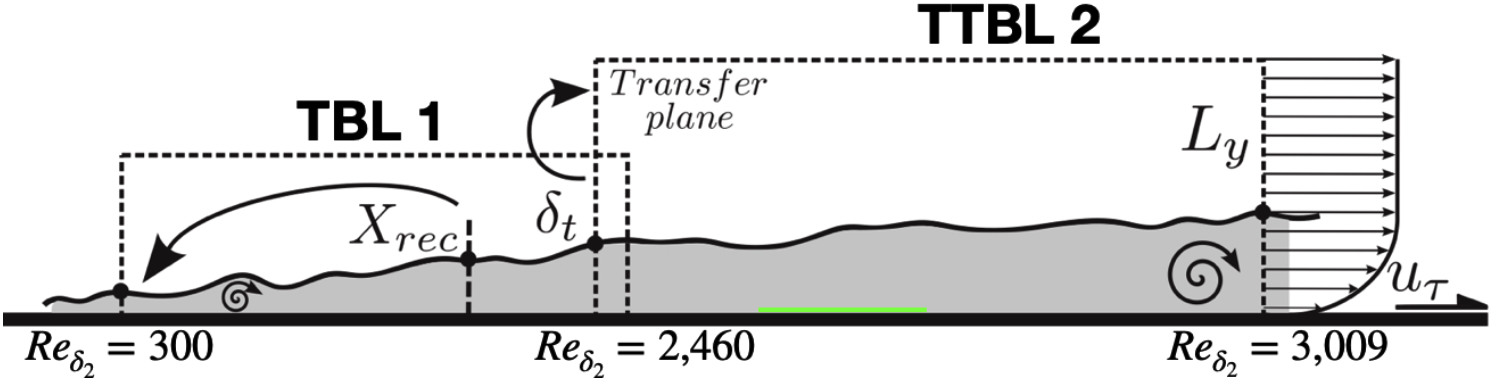}
	\caption{A dual \ac{tbl} \ac{dns} configuration. TTBL2 yields the \ac{ttbl} \ac{dns}, which is an analogy of a bushfire. TBL1 provides a proper inflow boundary condition to TTBL2. The thin green layer in TTBL2 is the biomass energy source.}
    \label{fig:Fig-6}
\end{figure}

A high-fidelity simulation was undertaken with a dual\ac{tbl} \ac{dns} as shown in Fig. \ref{fig:Fig-6} the computational domain characteristics for TBL1 and TTBL2 are given in Table \ref{tab:table-1} and  \ref{tab:table-2}, respectively.

\begin{table}[htbp]
\begin{tabular}{|p{1.5cm}|p{1.5cm}|p{1.5cm}|p{1.5cm}|p{1.5cm}|p{1.5cm}|}
\hline
\multicolumn{3}{|c|}{Computational Domain} & \multicolumn{3}{|c|}{Computational Grid} \\ 
\hline
$L_x$ & $L_y$ &  $L_z$ &  $N_x$ & $N_y$ &  $N_z$ \\  
\hline
$100\, \pi$ &  $30$ &  $33.4\, \pi$ &  $4,096$ &  $315$ &  $1,024$ \\  
\hline
\end{tabular}
\caption{Computational Domain for TBL1 in Fig. \ref{fig:Fig-6}.}
\label{tab:table-1}
\end{table}

\begin{figure}[htbp]
	\centering
	\includegraphics[width=\textwidth]{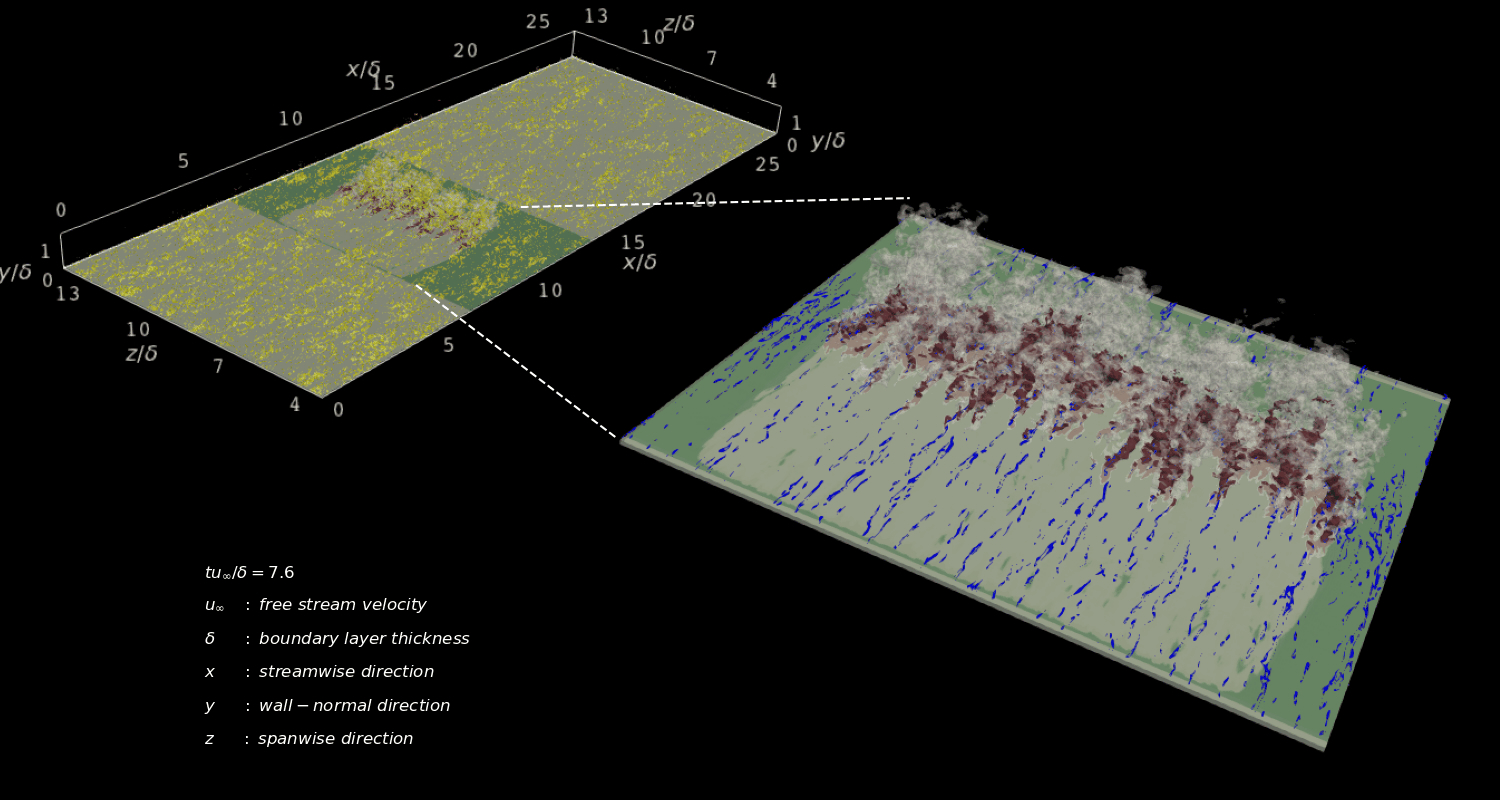}
	\caption{Visualisation of the \ac{ttbl} \ac{dns} with the green region indicating the biomass fuel, the yellow structures in the full computational domain being coherent vortical structures visualised using the second invariant of the velocity gradient tensor, while the blue streaks in the zoomed in view representing low velocity streak. The red colour in both indicates the bushfire front at the beginning of the bushfire. A full animation can be viewed at \cite{Li2022}.}
    \label{fig:Fig-7}
\end{figure}

\begin{figure}[htbp]
	\centering
	\includegraphics[width=\textwidth]{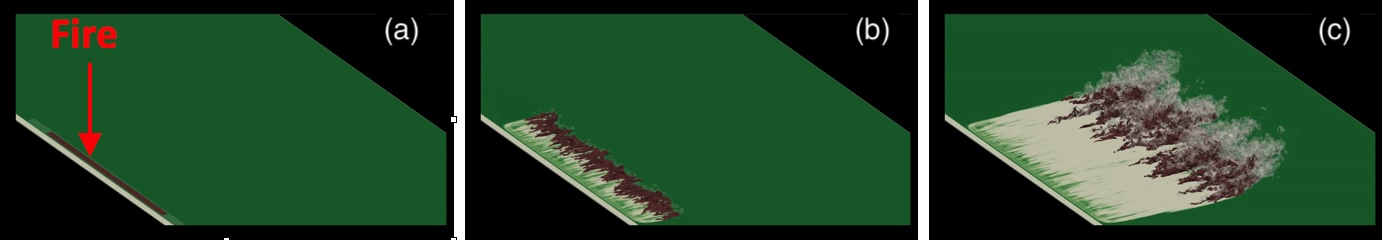}
	\caption{Some snapshots of the biomass burning evolution zoomed in from Fig. \ref{fig:Fig-7} (a) indicates the staring of the bushfire, (b) is a time after the bushfire has started, but with little smoke at this stage, (c) indicates the progress of the bushfire with the grey colour indicating fluid temperatures which are typical of smoke rather than fire.}
    \label{fig:Fig-8}
\end{figure}

The uniform biomass energy source is only located within TTBL2 and ranges in the streamwise direction starting at $\frac{L_x}{4}$ and ending at $\frac{L_x}{2}$, while spanning the entire width of the TTBL2 domain $L_z$. The uniform biomass height was set at $y^+ = 50$ referenced at $x = \frac{L_x}{4}$. The values used for the source term given by Eq. \ref{eq:SimplifiedSource}, which characterises the biomass heat release rate in Eq. \ref{eq:temp} are: $A = 50$,  $\alpha = 0.8$ and $\beta = 8$. The fire is started by instantaneously raising the temperature of the fluid above $\theta = 0.5$ over a small streamwise domain at $x = \frac{L_x}{4}$ spanning $0.8 L_z$ along the spanwise domain as shown in Fig. \ref{fig:Fig-7} and \ref{fig:Fig-8}.

\begin{table}[htbp]
\begin{tabular}{|p{1.5cm}|p{1.5cm}|p{1.5cm}|p{1.5cm}|p{1.5cm}|p{1.5cm}|}
\hline
\multicolumn{3}{|c|}{Computational Domain} & \multicolumn{3}{|c|}{Computational Grid} \\ 
\hline
$L_x$ & $L_y$ &  $L_z$ &  $N_x$ & $N_y$ &  $N_z$ \\  
\hline
$50\, \pi$ &  $40$ &  $33.4\, \pi$ &  $6,554$ &  $536$ &  $2,048$ \\  
\hline
\multicolumn{6}{|c|}{Computational domain normalised by boundary layer thickness at $L_x/4$} \\
\hline
\multicolumn{2}{|c|}{$L_x/\delta$} & \multicolumn{2}{|c|}{$L_y/\delta$} &  \multicolumn{2}{|c|}{$L_z/\delta$} \\  
\hline
\multicolumn{2}{|c|}{$27.7$} & \multicolumn{2}{|c|}{$7.1$} &  \multicolumn{2}{|c|}{$18.5$} \\  
\hline
\end{tabular}
\caption{Computational Domain for TTBL2 in Fig. \ref{fig:Fig-6}, $\delta$ is the \ac{tbl} thickness at the beginning of this computational domain.}
\label{tab:table-2}
\end{table}

Figure \ref{fig:Fig-7} shows the full computational domain of TTBL2 and a zoomed in version of the biomass domain. The yellow structures in the full computational domain represent coherent vortical structures visualised using the second invariant of the velocity gradient tensor \cite{Ooi99}, while the blue streaks in the zoomed in view representing low velocity streaks. The red colour in both indicates the bushfire front at the start of the bushfire.

Figure \ref{fig:Fig-8}, shows three snapshot along the bushfire simulation, beginning from the fire starting, evolving as a bushfire without too much of a smoke bloom and subsequently evolving into a bushfire with significant smoke reaching the upper parts of the \ac{ttbl}. A full animation of the bushfire analogy provided by this \ac{ttbl} \ac{dns} can be viewed at \cite{Li2022}. It is worth noting the qualitative similarities in the latter stages of the \ac{ttbl} \ac{dns} and the photograph of the Blue Mountain bushfire given in Fig. \ref{fig:Fig-1}.

\section{Conclusion}

This paper has proposes a \ac{sciml} methodology that utilises physics-informed \ac{ml} based on \ac{dl} - \ac{pidl} to assimilate remotely sensed temperature data and the approach to its \ac{uq}. 
The development, testing and \ac{uq} is entirely reliant on the \ac{ttbl} \ac{dns}, which is a canonical analogy of a bushfire, to provide training data and the {\em “ground truth”} for \ac{uq}. Ultimately in the field the \ac{pidl} will learn from remotely sensed 3D temperature data gathered from a variety of sources ranging from low orbit satellites to fixed wing aircraft and UAVs or drones fitted with infrared sensors and be able to predict the thermal turbulent boundary layer associated with a bushfire, as well as predict its dynamics and fire front spreading rate.
A realistic heat source rate model has been developed based on approaches which have their roots in the classical developments of theoretical combustion. This model has been employed to undertake an initial canonical bushfire analogy employing a uniform biomass distribution. The resulting \ac{ttbl} \ac{dns} has shown close qualitative aspects with images of the recent Blue Mountains Bushfire in Australia.

\section*{Acknowledgements}
This research was supported by a NCI Australasian Leadership Computing Grant and a Monash Data Futures Institute Seed Grant. The authors also gratefully acknowledge the NCMAS HPC allocation of NCI supported by the Australian Federal Government and of Pawsey also supported by the Australian Federal Government and the State Government of Western Australia.

\bibliographystyle{sn-bibliography}
\bibliography{references}

\end{document}